\documentclass[twocolumn,showpacs,preprintnumbers,amsmath,amssymb,superscriptaddress,epsfig,pra]{revtex4}

\usepackage{graphicx}
\usepackage{dcolumn}
\usepackage{bm}
\usepackage{color}

\def\fig_width{3. in} 

\begin{document}

\title{Transverse laser cooling of a thermal atomic beam of dysprosium}

\author{N. Leefer}
\email{naleefer@berkeley.edu}
\affiliation{Department of Physics, University of California at Berkeley, Berkeley, CA 94720-7300, USA}

\author{A. Cing{\"{o}}z\footnote{currently at: JILA, National Institute of Standards and Technology and University of Colorado at Boulder, Boulder, CA 80309-0440, USA}}
\affiliation{Department of Physics, University of California at Berkeley, Berkeley, CA 94720-7300, USA}

\author{B. Gerber-Siff}
\affiliation{Swarthmore College, Swarthmore, PA 19081, USA}

\author{Arijit Sharma}
\affiliation{Raman Research Institute, Sadashivanagar, Bangalore 560080, India}

\author{J. R. Torgerson}
\affiliation{Physics Division, Los Alamos National Laboratory, Los Alamos, NM 87545, USA}

\author{D. Budker}
\email{budker@berkeley.edu}
\affiliation{Department of Physics, University of California at Berkeley, Berkeley, CA 94720-7300, USA} \affiliation{Nuclear Science Division, Lawrence Berkeley National Laboratory, Berkeley, California 94720, USA}

\date{\today}



\begin{abstract}
A thermal atomic beam of dysprosium (Dy) atoms is cooled using the $4f^{10}6s^2\,(J=8)\,\rightarrow\,4f^{10}6s6p\,(J=9)$ transition at 421 nm. The cooling is done via a standing light wave orthogonal to the atomic beam. Efficient transverse cooling to the Doppler limit is demonstrated for all observable isotopes of dysprosium. Branching ratios to metastable states are demonstrated to be $<5\times10^{-4}$. A scheme for enhancement of the nonzero-nuclear-spin-isotope cooling, as well as a method for direct identification of possible trap states, is proposed.
\end{abstract}


\maketitle

\section{Introduction}

Laser cooling and trapping of atoms has been a prolific area of research, with applications to the study of novel quantum phases~\cite{Lewenstein2000,Ketterle2008}, matter-wave interferometry~\cite{Muller2008}, quantum information~\cite{Lewenstein2008}, and precision measurements and tests of fundamental physics~\cite{Veronesi2003}. Atoms amenable to direct laser cooling have recently expanded outside of the alkalis and noble gases to include elements such as chromium, indium, ytterbium and mercury~\cite{Tao2008,Kim2009,Yoon2003,Katori2008}. The lanthanides are of particular interest. The generally large ground-state magnetic moments of these elements present opportunities for new studies in areas such as deterministic single atom sources~\cite{McClelland2003}, degenerate dipolar Fermi gases~\cite{Lev2009}, and quantum computing~\cite{McClelland2005}. While the complex electronic structures and numerous possible decay channels for excited states do not make most lanthanides obvious candidates for laser cooling, some progress has been made. In erbium, the large magnetic moment allows for magnetic confinement in the trap region while excited states recycle to the ground state, permitting laser cooling to low temperatures~\cite{Hanssen2006}. A successful effort to form a Dy magneto-optical trap (MOT) has demonstrated that the same principle works for dysprosium~\cite{Lev22009}.

In this work, we explore transverse laser cooling of dysprosium (Dy, Z=66) in an atomic beam. In addition to sharing the largest ground-state magnetic moment in the periodic table with terbium ($\mu \sim 10\,\mu_B$), Dy contains a pair of nearly degenerate, opposite-parity states the have been investigated in detail~\cite{BudkerThesis}. These levels have recently been used in a search for a variation of the fine-structure constant that does not rely on optical-frequency combs~\cite{Nguyen2004,Cingoz2007}. The implementation of laser cooling in this experiment as a control over atomic-beam parameters is a crucial aspect of systematics management~\cite{ArmanThesis}. Previously our group reported on the manipulation of a thermal beam of Dy atoms using a bright, visible transition in Dy at 421 nm~\cite{Leefer2008} (which was the first report of laser cooling in Dy) and the spectroscopy of this transition~\cite{Leefer2009}. In this work, we demonstrate successful transverse laser cooling of a thermal beam for all isotopes of dysprosium without repumping, propose a method for the simultaneous cooling and optical pumping within the hyperfine manifold of the odd-neutron-number isotopes with non-zero nuclear spin, and show that the maximum possible branching ratio of the cooling transition to metastable states is less than $5\times10^{-4}$.

\begin{table}[b]
\caption{Parameters of the cooling transition }
\center
\begin{tabular*}{0.47\textwidth}{@{\extracolsep{\fill}} l  l  r }
\hline
Parameter & & Value\\
\hline\hline
Wavelength & $\lambda_c$& 421.291 nm\\
Lifetime Exc. State~\cite{Curry} & $\tau$ & 4.8 ns\\
Recoil Velocity & $v_r=\hbar k/M$ &  0.6 cm/s\\
Doppler Limit & $v_d=\sqrt{\hbar /2 M \tau}$ & 19.8 cm/s\\
Saturation Intensity & $I_{sat} = h c \pi/2 \tau \lambda_c^3$ & 60 mW/cm$^2$\\
\hline
\end{tabular*}
\label{table:transition}
\end{table}

\section{Experiment}

\subsection{Dysprosium}

Dysprosium has seven stable isotopes, ranging from $A=156\rightarrow164$, including two isotopes, $^{161,163}$Dy, with nuclear spin $I=5/2$. Cooling is accomplished using the $4f^{10}6s^2\,(J=8)\rightarrow4f^{10}6s6p\,(J=9)$ transition at 421 nm~\cite{Leefer2009}. Cooling of the odd-neutron-number isotopes was accomplished using the only closed transition in the hyperfine manifold from $F=10.5\rightarrow\,F'=11.5$. The relevant cooling parameters of this system are given in Table~\ref{table:transition}.

\subsection{Apparatus}

This work was carried out using a beam source previously used in a search for a variation of the fine-structure constant, a search for parity nonconservation (PNC), and rf-spectroscopy of nearly degenerate levels in Dy~\cite{Cingoz2007, Ferrell2007, Nguyen1997, BudkerThesis}. A detailed description of the atomic-beam source is given in Ref.~\cite{Nguyen1997}. The beam is produced by an effusive oven with a multislit nozzle array operating at $\simeq$~1500~K. The oven consists of a molybdenum tube containing dysprosium metal, and is surrounded by resistive heaters made from tantalum wire enclosed inside alumina ceramic tubes. The atoms in the beam have a mean longitudinal velocity of $\simeq$~3$\times 10^4$~cm/s. The 1D transverse velocity profile was measured via fluorescence spectroscopy to be Gaussian (Eqn.~\ref{eqn:gauss}), with $\sigma_v = 1.3\times10^3$~cm/s.

\begin{figure}[t]
\begin{center}
\includegraphics[width=8.4cm]{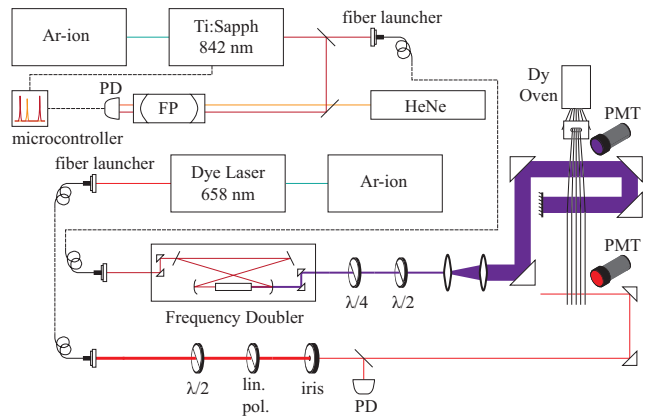}
\caption{(Color online) Simplified layout of the experimental system. Light from two Ar-ion lasers pumped a Coherent 899 Ti:Sapphire and Coherent 699 dye laser respectively. The output of the Ti:Sapph laser was split, with ~10\% of power going to diagnostics and the frequency stabilization system. The rest was fiber coupled to the frequency doubling system, the output of which was expanded and polarized before entering the vacuum chamber. Light from the dye laser was fiber coupled to the beam apparatus, where it passed through polarization optics and an iris diaphragm before entering the chamber.}
\label{fig:schematic}
\end{center}
\end{figure}

To generate the 421-nm light, 12~W of Ar-ion laser light (Coherent Innova 400) was used to pump a Ti:Sapphire ring laser (Coherent 899), producing up to 650 mW of 842-nm light.  After diagnostics and passing through an optical fiber, approximately 300 mW of 842-nm light entered a resonant bow-tie cavity containing a 1$\times$2$\times$10 mm$^3$ periodically poled potassium titanyl phosphate (PPKTP) crystal (Raicol Crystals Ltd.) located at the 22-$\mu$m beam waist of the cavity. The output power of the frequency doubler was consistently 50-60 mW. Outputs powers of $>$100 mW were observed, but at powers above 60 mW, gray tracking in the crystal caused rapid degradation of the conversion efficiency~\cite{Marnier1996}. It was determined experimentally that laser-cooling did not noticeably improve above $\sim$~30~mW of light power, so the cavity was operated at sufficiently lower power to eliminate noticeable degradation.  The 421-nm light passed through a quarter- and a half-wave plate to adjust the polarization to vertical and a 4x cylindrical telescope to create a $0.5$~mm~$\times$~$1.5$~mm rectangular beam profile . The 421-nm beam entered the atomic-beam apparatus and intersected the atoms at a ninety-degree angle in two regions (see Fig.~\ref{fig:schematic}). The total length of the cooling region was approximately 3 cm, giving a total interaction time for an atom in the beam on the order of 100~$\mu$s. Assuming a cycle time of $4\tau \simeq20$~ns for a saturation parameter of $\sim1$~\cite{Metcalf}, this allows for $5\times10^3$ cycles.

To reduce long term drifts of the laser frequency, the 421-nm light was scanned over the desired resonance while monitoring fluorescence with a photomultiplier tube (PMT) oriented at 90 degrees with respect to the laser and atomic beams. Once at the desired frequency, the Ti:Sapph laser was stabilized using a scanning transfer-cavity lock~\cite{Sackett2005}. In this control system, light at 842 nm from the Ti:Sapph laser is injected into a continuously scanning Fabry-Perot (FP) cavity along with light from a metrological HeNe (Spectra-Physics 117A) with a frequency drift rate of $<0.5$~MHz/hr. The transmitted light is detected with a photodiode and sent to a peak-detection and timing circuit that measures the time delay between HeNe and Ti:Sapph transmission peaks. These time data are sent to an IsoPod microcontroller (New Micros, Inc.) which then adjusts the frequency of the Ti:Sapph to keep the time delay constant. The frequency drift of the Ti:Sapph stabilized this way was measured to be $<2$~MHz/hr~\cite{ArmanThesis}.

After passing through the cooling region, the atoms traversed $\sim$~15~cm to the back of the apparatus for detection. The Doppler profile of the atomic beam was characterized by scanning a probe-laser frequency over the $4f^{10}6s^2\,(J=8)\,\rightarrow\,4f^{9}5d6s^2\,(J=7)$ transition at 658 nm. For a Gaussian velocity distribution described by

\begin{equation}\label{eqn:gauss}
P(v) = \frac{1}{\sqrt{2\pi}\sigma_v} e^{-\frac{v^2}{2\sigma_v^2}}
\end{equation}
with characteristic half-width 1/e$^2$ velocity $2\sigma_v$, the pure Doppler spectral profile is also Gaussian with the relationship $\sigma_f = k \sigma_v/2\pi$, where $k = 2 \pi/\lambda$.  The Doppler cooling limit, $v_d = 19.8$~cm/s (see Table~\ref{table:transition}), corresponds to a spectral profile with a width $\sigma_{v_D}=300$~kHz at 658 nm. Thus, the 658-nm transition is an ideal probe of the atoms' Doppler profile, as the natural width of the transition, $120$~kHz~\cite{Curry}, is considerably smaller than the Doppler limited width. The sensitivity in our experiment was ultimately limited by the probe-laser linewidth; effects such as magnetic-field broadening and power broadening were controlled and maintained at a smaller level.

The 658-nm light was produced using a Coherent 699 dye laser pumped by 5.5 W of Ar-ion laser light. After passing through an optical fiber and diagnostics, $\simeq$~50~$\mu$W of vertically polarized light in a 4-mm diameter beam entered the apparatus and intersected the atomic beam at a ninety-degree angle. Fluorescence from the probe transition was detected using a PMT oriented perpendicular to the laser and atomic beams, with a 660-nm central wavelength interference filter (10-nm bandwidth) on the detection port. The dye-laser frequency was scanned continuously over the transition, while the fluorescence signal from the PMT and the signal from a reference photodiode (PD) were simultaneously acquired using an Agilent DSO5014A oscilloscope. Normalizing the fluorescence signal with the reference PD signal ensured that fluctuations in the probe laser power did not contribute to features in the Doppler profile.

\subsection{Results\label{sec:results}}

\begin{figure}[t]
\begin{center}
\includegraphics[width=8.4cm]{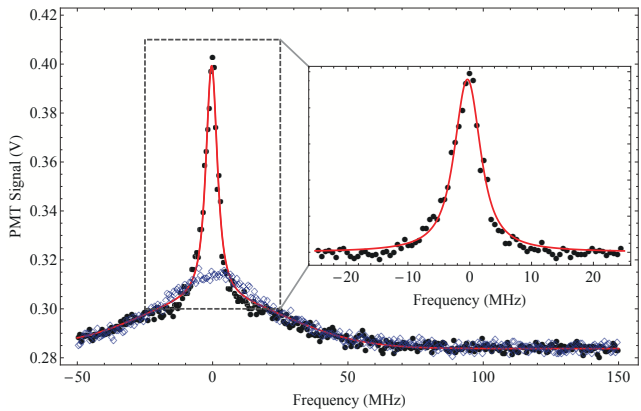}
\caption{(Color online) Example probe spectrum of the $^{164}$Dy 658-nm transition with 40 mW of cooling light. The uncooled Doppler profile is shown by the unfilled diamonds. Optical pumping in the cooling region to magnetic sublevels with larger Clebsch-Gordan coefficients accounts for the non-conservation of area under the curve. The solid red line is a nonlinear least-squares fit to the data. The inset shows the narrow feature with the broad background subtracted and a single Voigt profile fit to the remaining peak.}
\label{fig:cooling}
\end{center}
\end{figure}

A scan over the 658-nm $^{164}$Dy resonance is shown in Fig.~\ref{fig:cooling}. The broad uncooled Doppler profile has a Gaussian profile with characteristic width, $\sigma_{f_i} = 20$~MHz, corresponding to $\sigma_{v_i}=1.3\times10^3$ cm/s. At low temperatures the measured lineshape is expected to be described by a Voigt profile, a convolution of the Doppler spectrum of the atoms and the Lorentzian lineshape of the probe laser, given by

 \begin{align}
 V(f)=&\\
 \int_{-\infty}^\infty\frac{1}{\sqrt{2\pi}\sigma_{f_c}}&\mathrm{exp}\left(-\frac{(f-f')^2}{2\sigma_{f_c}^2}\right)\frac{1}{\pi}\frac{\gamma_L/2}{f'^2+\gamma_L^2/4}df',\notag
 \end{align}
where $\sigma_{f_c}$ and $\gamma_L$ are the Gaussian width of the cooled thermal distribution and the full-width half-maximum (FWHM) of the Lorentzian describing the laser spectrum, respectively. For the best cooling, a good fit was achieved using a combination of a Gaussian and a Voigt profile: a Gaussian profile for the uncooled background and a Voigt profile for the narrow feature (see inset in Fig.~\ref{fig:cooling}). The uncooled background is due to atoms outside the capture range of the cooling laser. The values for $\sigma_{f_c}$ and $\gamma_L$ obtained from a least-squares fit are $0.8(5)$~MHz and $4.2(7)$~MHz, respectively. The value for $\sigma_{f_c}$ gives $\sigma_{v_c} = 53(33)$~cm/s, which is consistent with the expected Doppler cooling limit of 19.8 cm/s (Table~\ref{table:transition}). The value for $\gamma_L$ is also consistent with a separate measurement of the laser linewidth (Sec.~\ref{sec:broad}).

\begin{figure}[t]
\begin{center}
\includegraphics[width=8.4cm]{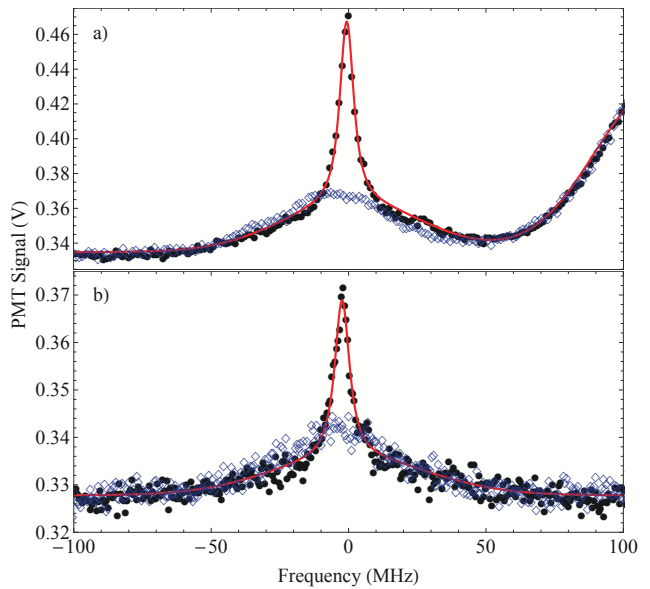}
\caption{(Color online) Probe spectra of the a) $^{163}$Dy and  b) $^{161}$Dy 658-nm, $F=10.5\rightarrow9.5$ transitions with $\sim40$~mW of cooling light. The unfilled diamonds show the uncooled Doppler profile, and the solid red line is a nonlinear least-squares fit with a Voigt profile used for the narrow feature. The slope on the right side of a) is from the edge of the $^{162}$Dy transition.}
\label{fig:oddisotopes}
\end{center}
\end{figure}

Laser cooling of $^{161}$Dy and $^{163}$Dy, both of which have hyperfine structure, was accomplished by tuning the cooling laser onto the closed $F=10.5\rightarrow11.5$ transition, where $F$ is the total angular momentum (Fig.~\ref{fig:oddisotopes}). The Gaussian widths of the Voigt profiles fit to the narrow features of $^{163}$Dy and $^{161}$Dy spectra are $0.7(7)$~MHz and $0.9(4)$~MHz, respectively, and agree with the Doppler cooling limit.

The number of cycles can be estimated from the initial velocity, ($\sigma_{v_i} = 1300$~cm/s), the final velocity, and the recoil velocity given in Table~\ref{table:transition} as $N\approx (\sigma_{v_i}-\sigma_{v_f})/v_r\approx\sigma_{v_i}/v_r\sim$~2000. This allows us to constrain the branching ratio as $1/N<5\times10^{-4}$. This result is in agreement with a more sensitive measurement of branching ratios that has been performed in a MOT, yielding $7\times10^{-6}$~\cite{Lev22009}.

\subsection{Broadening Mechanisms\label{sec:broad}}

\begin{figure}[t]
\begin{center}
\includegraphics[width=8.4cm]{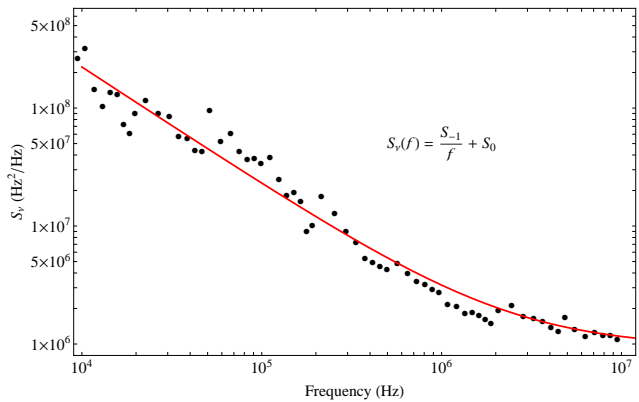}
\caption{(Color online) Power spectral density of frequency fluctuations of the dye laser used to observe the 1D temperature of the atomic beam. The solid red line is a least-squares fit to a function of the form given by Eqn.~\ref{eqn:PSD}}
\label{fig:linewidth}
\end{center}
\end{figure}

Our measurement of the laser cooled transverse temperature of the atomic beam was ultimately limited by the linewidth of the dye laser. The linewidth was measured by coupling the 658-nm light into a FP cavity stabilized to a metrological HeNe (Spectra-Physics 117A). The dye laser was tuned to the side of a transmission peak, where the slope is linear and the fluctuations in the transmitted power were recorded with a spectrum analyzer. The known free-spectral range (FSR) and measured finesse of the cavity allowed us to to convert the voltage power spectral density (PSD) of the fluctuations to the frequency PSD [$S_\nu (f)$]. The result of this measurement is shown in Fig.~\ref{fig:linewidth}. A function of the form,

\begin{equation}\label{eqn:PSD}
S_\nu(f) =  \frac{S_{-1}}{f} + S_0
\end{equation}
was fit to the data, where $S_{-1}$ and $S_0$ are the $1/f$- and white-noise contributions, respectively~\cite{Riehle}. The least-squares fit values for $S_{-1}$ and $S_0$ are $2.2(1)\times10^{12}$~Hz and $9.4(4)\times10^5$~Hz respectively. Interpretation of the shape of the laser linewidth from this data is non-trivial, but the FWHM can be estimated by solving the equation~\cite{Hall}

\begin{equation}\label{eqn:FWHM}
1\,\mathrm{rad}^2 \approx \int_{f_{3dB}}^\infty \frac{S_\nu(f)}{f^2}df,
\end{equation}
where $2f_{3dB}$ is the FWHM of the laser lineshape. Solving Eqn.~\ref{eqn:FWHM} for $f_{3dB}$ we find a FWHM of 3.2(5)~MHz, in agreement with the value obtained from a Voigt profile fit (Sec.~\ref{sec:results}).

The most important remaining broadening mechanisms of the probe transition linewidth are magnetic-field broadening and power broadening. The magnetic-field broadening is given by

\begin{equation}
\Delta \nu_B =| (m_{J_u}g_u - m_{J_l}g_l)| \mu_B B,
\end{equation}
where $g_u=1.26$ and $g_l=1.24$ are the $g$-factors of the upper and lower energy levels and $m_{J_u}$ and $m_{J_l}$ are the magnetic sublevels involved. The worst-case scenario of driving $m\rightarrow m\pm1$ transitions gives a maximum broadening of $\sim$~1.9~MHz/G. This effect was minimized by coils installed around the probe region, which reduce the background magnetic field to $\sim$~1~mG.

The other broadening mechanism of concern is power broadening. The saturation intensity, $I_S$, of the probe transition is approximately 50~$\mu$W/cm$^2$, and the corresponding power broadened linewidth can be calculated using

\begin{equation}
\gamma_P = \gamma_0 \sqrt{1+\kappa},
\end{equation}
where $\gamma_0$ is the natural linewidth and $\kappa = I/I_S$ is the saturation parameter. With an approximate beam radius of 2 mm, an input power of 50~$\mu$W gives $\kappa\simeq8$ and $\gamma_P\simeq360$~kHz, well below the linewidth of our laser. In addition, data were taken as a function of probe laser power to confirm the absence of measurable power broadening.

\subsection{Hyperfine Structure\label{sec:odd}}

\begin{figure}[t]
\begin{center}
\includegraphics[width=8.4cm]{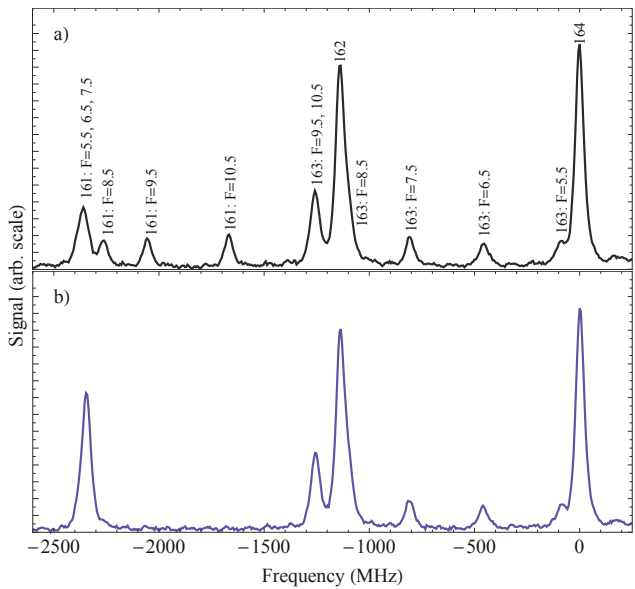}
\caption{(Color online) Example of the optical pumping dynamics observed. a) A typical spectrum of the 658-nm transition. For the odd-neutron number isotopes, the strongest $\Delta F = -1$ transitions are labeled with the ground state angular momentum. b) The same spectrum with the 421-nm light tuned into the vicinity of the $\Delta F =0$ transitions. See Section~\ref{sec:odd} for details.}
\label{fig:opticalpumping}
\end{center}
\end{figure}

The dense structure of the 421-nm cooling transition gives rise to interesting optical pumping dynamics (spectroscopic information on the 421-nm spectrum can be found in Ref.~\cite{Leefer2009}). In $^{161}$Dy, all of the $\Delta F = 0$ transitions are unresolved due to the broad natural linewidth and Doppler width. Because the cooling transition is a $\Delta J = +1$ transition in the uncoupled angular-momentum basis ($J$ is the total electronic angular momentum), driving all of the $\Delta F = 0$ transitions makes the ground state with the smallest value of $F$ a dark state. In Fig.~\ref{fig:opticalpumping}, the 421-nm laser has been tuned onto the cluster of $\Delta F = 0$ transitions while the probe laser is swept over the entire 658-nm spectrum, demonstrating optical pumping of the atoms into the $F=5.5$ ground state.

This suggests a convenient method for simultaneous cooling and optical pumping of $^{161}$Dy. The 421-nm carrier frequency can be tuned to drive the $F=5.5\rightarrow6.5$ transition, while a single modulation at $\sim1$~GHz can be used to generate a sideband in the vicinity of the $\Delta F = 0$ transitions. The sideband will optically pump atoms towards the $F=5.5$ ground state, while the carrier cools these atoms on the $F=5.5\rightarrow6.5$ transition. This has the potential to eliminate the multiple sidebands that would normally be required to cool the entire hyperfine manifold, as described in the next section.

\begin{figure}[t]
\begin{center}
\includegraphics[width=8.4cm]{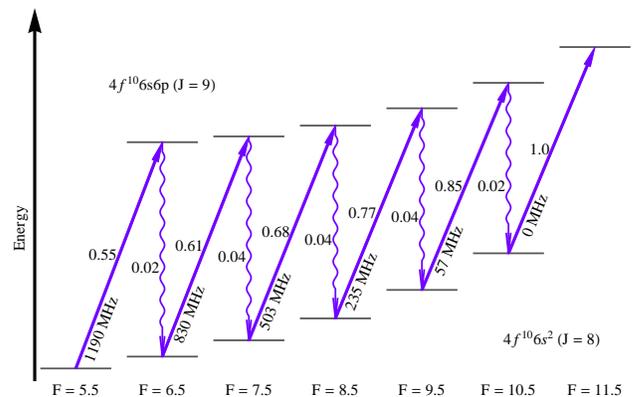}
\caption{(Color online) Diagram of the $^{163}$Dy hyperfine structure. The relative transition strengths $||d_F||^2/||d_J||^2$ are shown alongside transition arrows, where $||d||$ is the reduced dipole matrix element normalized so that the $F=10.5\rightarrow 11.5$ strength is equal to 1. The frequencies of the $F\rightarrow F+1$ transitions relative to the $F=10.5\rightarrow 11.5$ transition were calculated from known hyperfine coefficients~\cite{Childs,Leefer2009}. The $\Delta F = -\Delta J$ transitions are negligibly weak and are not shown.}
\label{fig:hyperfinemanifold}
\end{center}
\end{figure}

\section{Future Work}

The extension of laser cooling to include all observable isotopes of Dy has been demonstrated, with cooling of the non-zero-nuclear-spin isotopes, $^{163}$Dy and $^{161}$Dy, accomplished using the closed $F=10.5\rightarrow11.5$ transition. To improve the efficiency of cooling the odd-neutron number isotopes, we propose a scheme for cooling on the closed $F=10.5\rightarrow11.5$ transition while simultaneously transferring the population to the $F=10.5$ state from some or all of the other hyperfine ground states. A diagram of the hyperfine manifold for $^{163}$Dy is shown in Fig.~\ref{fig:hyperfinemanifold}.

This can be accomplished by passing the 421-nm light through a running-wave (non-resonant) electrooptic modulator~\cite{Brewer}. The desired modulation frequencies can all be applied to the same crystal to generate sidebands at the appropriate frequencies. Atoms being driven on the $F\rightarrow F+1$ transitions will eventually be pumped to the $F=10.5$ ground state where they will continue to cycle. The effectiveness of this method will be limited by the power available to drive the modulator, however methods such as those described in Sec.~\ref{sec:odd} may allow us to circumvent these limitations.

Given the complex electronic structure of the dysprosium system, the existence of a closed optical transition is unlikely. A theoretical calculation was performed that estimated the total branching ratio to the most likely trap states at less than the $5\times10^{-5}$ level~\cite{DzubaPersonal} and a measurement in a MOT obtained a value of $7\times10^{-6}$~\cite{Lev22009}. Detailed knowledge of the dominant trap states will be important for optimization of MOTs and future work with laser cooled and trapped Dy.

\begin{figure}[t]
\begin{center}
\includegraphics[width=8.4cm]{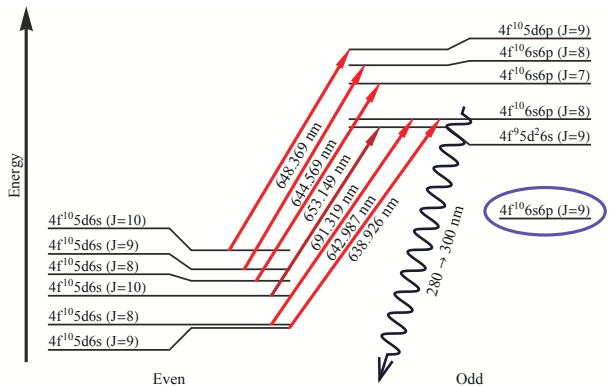}
\caption{Partial energy level diagram showing possible repump wavelengths from the likely trap states. The circled state is the upper level of the cooling transition. The trap states and corresponding repump wavelengths were identified in the NIST Atomic Spectroscopy database.}
\label{fig:repumpstates}
\end{center}
\end{figure}

Our current set-up is insensitive to trap states for which the time of return to the ground state is less than the relatively long time-of-flight, $\tau_{tof}\sim$~500$~\mu$s, between the cooling and probe regions. We have developed a procedure to directly identify trap states. The 421-nm light will be tuned on resonance with the $^{164}$Dy transition while light from a pulsed dye laser operating between 600 and 700 nm will be superimposed over the fluorescence region. A PMT with a UV short-pass filter on the viewport will be oriented perpendicular to the atomic beam and the cooling laser while the dye laser is scanned from 600 to 700 nm. An increase in the fluorescence signal from the PMT will be taken to indicate a transition from the trap state to a highly excited odd-parity state, the fluorescence from the ground state decay channel of this state being less than $300$ nm. A diagram of likely repump wavelengths is presented in Fig.~\ref{fig:repumpstates}.

\section{Discussion}

We have presented results on transverse cooling of a thermal dysprosium beam. A final laser linewidth limited spectrum of a narrow probe transition in the cooled beam consistent with cooling to the Doppler limit was obtained. Cooling has been demonstrated for all detectable isotopes of Dy. A separate measurement of the laser linewidth was consistent with the hypothesis that the laser linewidth is the limiting factor in measuring the Doppler spectrum. An attempt was made at sub-Doppler cooling, but no narrowing of the line was observed due to the large probe laser linewidth. The remaining dominant broadening mechanisms of the probing method, background magnetic-field and power broadening, were estimated and controlled for. A plan for increasing the efficiency of cooling the odd-neutron number isotopes of Dy has been presented. The branching ratios of the cooling transition have been confirmed to be $<5\times10^{-4}$ and a procedure for direct identification of the dominant trap states has been outlined.

Future work includes using a femtosecond frequency comb as a probe of the Doppler profile of the atomic beam. The narrow width of the comb teeth should be sufficient to directly observe Doppler and sub-Doppler cooling. The ultimate goal of this work is the extension to 2D transverse cooling and longitudinal slowing in an experiment dedicated to a search for variation of the fine-structure constant.

\acknowledgements{The authors would like to acknowledge the contributions of V. Flambaum and V. Dzuba for the theoretical calculations of branching ratios and H. M\"{u}ller for comments on the manuscript. We would like to thank A. Skliar of Raicol Crystals Ltd. for extensive technical advice regarding the PPKTP crystal. This work has been supported by the Foundational Questions Institute (fqxi.org), the UC Berkeley Committee on Research and NSF/DST Grant No. PHY-0425916 for U.S.-India cooperative research.}

\bibliography{DyCoolingBib}

\end{document}